\documentclass[12pt]{JHEP3}
\usepackage{graphicx}
\usepackage{amsmath,amssymb,amsfonts}

\newcommand{\beq}{\begin{equation}}
\newcommand{\eeq}{\end{equation}}
\newcommand{\beqn}{\begin{eqnarray}}
\newcommand{\eeqn}{\end{eqnarray}}

\newcommand{\comment}[1]{}
\newcommand{\nn}{\nonumber}
\newcommand{\ZZ}{\mathbb{Z}}

\title{\bf Gravitational Duality Transformations on (A)dS$_4$ }
\author{\bf Robert G. Leigh\\Department of Physics\\
University of Illinois at Urbana-Champaign\\
1110 West Green Street, Urbana, IL 61801-3080, USA\\
Email: \email{rgleigh@uiuc.edu} }

\author{Anastasios C. Petkou\\
Department of Physics\\
University of Crete\\
Heraklion 71003, Greece\\
Email: \email{petkou@physics.uoc.gr}}
\preprint{hep-th/yymmnnn\\ILL-(TH)-07-02}

\abstract{We discuss the implementation of electric-magnetic duality transformations in four-dimensional gravity linearized around Minkowski or (A)dS$_4$ backgrounds. In the presence of a cosmological constant duality generically modifies the Hamiltonian, nevertheless the bulk dynamics is unchanged. We pay particular attention to the boundary terms generated by the duality transformations and discuss their implications for holography. 
}
\begin{document}

\section{Introduction and Summary}

Duality has played an important role in our understanding of Yang-Mill theories and it is believed that it will play an important role also in gravity and in higher-spin gauge theories.  Indeed, although it is less clear what could be the implications of duality for theories whose quantum versions are still unknown, gravity and higher-spin gauge theories\footnote{For reviews of higher-spin theories see e.g. \cite{HS_reviews}.}   are intimately connected to a quantum string theory where certainly duality plays a crucial role.

The recent advent of holography raises some intriguing questions for duality. For example one may wonder what is the holographic image of a duality invariant spectrum, a duality transformation or a possible quantization condition that usually duality implies for charges. Some of these issues were raised by Witten in \cite{Witten_sl2z} where it was argued that the standard electric-magnetic duality of a $U(1)$ gauge theory on AdS$_4$ is responsible for a ``natural" $SL(2,\ZZ)$ action on current two-point functions in three-dimensional CFTs.\footnote{See \cite{various_sl2z} and \cite{Sebas} for more recent works.} Shortly afterwards it was shown in \cite{LP1} that such an $SL(2,\ZZ)$ action is intimately related to certain ``double-trace" deformations in the boundary, assuming suitable large-N limits and existence of non-trivial  fixed points. The latter assumptions are strengthened by the fact that there exist models (e.g., see \cite{Hands} and references therein) which  exhibit the required behavior. In particular, it was shown in \cite{LP1} that certain "double-trace" deformations induce an $SL(2,\ZZ)$ action on two-point functions of higher-spin (i.e. spin $s\geq 2$) currents. This has led to the {\it Duality Conjecture} of \cite{LP1}: {\it linearized higher-spin theories on AdS$_4$ spaces possess a generalization of electric-magnetic duality whose holographic image is the natural $SL(2,\ZZ)$ action on boundary two-point functions}. 

Surprisingly, even the duality for linearized spin-2 gauge fields (linearized gravity) was not widely known by the time of this conjecture.\footnote{An interesting formulation of  first order duality for linearized gravity around flat space was presented in \cite{West}.} Second order linearized gravitational duality was discussed among other in \cite{Curtright,Nieto, Hull, Bekaert, Boulanger}. More recently, the duality properties of linearized gravity around flat space were studied in \cite{HT1} and were further discussed in \cite{Deser}. The duality of linearized gravity around dS$_4$ was later studied in \cite{Julia}.

In this note we present our calculations  regarding the duality properties of gravity in the presence of a cosmological constant.  Having in mind applications to higher-spin gauge theories we use forms and work in the first order formalism where duality is also manifested at the level of the action \cite{DT}. Moreover, the first order formalism is relevant for applications of duality to holography, since the correlation functions of the boundary theory are essentially determined by the bulk canonical momenta (see e.g. \cite{PapSkend}).

Our aim in this work is to formulate linearized first order gravity using  suitable "electric" and "magnetic" variables, in close analogy with electromagnetism. We find that this is possible {\it only} when the background geometry is Minkowski or (A)dS$_4$. Then we implement the standard electric-magnetic duality rotations. We find that, up to "boundary" terms,  the linearized Hamiltonian changes by terms that do not alter the bulk dynamics i.e. do not alter the second order bulk equations of motion.  Moreover, the duality rotation interchanges the (linearized) constraints with the (linearized) Bianchi identitites. The "boundary" terms have important holographic consequences since they correspond to marginal  "double-trace" deformations \cite{LP1} that induce the boundary $SL(2,\ZZ)$ action. 
In the Appendix we exhibit a modified duality rotation that leaves the bulk Hamiltonian invariant and induces "boundary" terms that correspond to relevant deformations as in \cite{LP1}.

\section{Action and Hamiltonian}

Having in mind the extension of our results to higher-spin gauge theorieswe start from the MacDowell-Mansouri form \cite{MM} of the gravitational action\footnote{We note $I=-16\pi G_N S$, where $S$ is the usually normalized gravitational action.}
\begin{equation} \label{eq:MMaction}
I_{MM}=\frac{1}{2\Lambda}\int_M \epsilon_{abcd} \left( R^{ab}\wedge R^{cd}
+2\Lambda e^a\wedge e^b\wedge R^{cd}+\Lambda^2 e^a\wedge e^b\wedge e^c\wedge e^d\right)\,,
\end{equation}
where $a,b,...$ are Lorentz indices.
In this formalism, the vierbein $e^a$ and the spin  connection ${\omega^a}_b$ are initially thought of as independent variables.  The curvature 2-form is
\[
{R^a}_b = d{\omega^a}_b+{\omega^a}_c\wedge {\omega^c}_b=\frac12 {R^a}_{bcd}e^c\wedge e^d
.\] 
Varying the action with respect to $e^a$ and ${\omega^a}_b$, we find
\beqn
\label{GReom1}
R^{ab}+\Lambda e^a\wedge e^b =0\,,\\
\label{GReom2}
T^a =de^a+{\omega^a}_b\wedge e^b=0\,.
\eeqn
The relation to gravity is established via the vanishing torsion equation (\ref{GReom2}), which relates $e$ and $\omega$ in the familiar way. The above equations are equivalent to the Einstein equation in metric variables
\beq
 R_{\mu\nu}-\frac12 Rg_{\mu\nu}=+3\Lambda g_{\mu\nu}\,.
\eeq
and the scalar curvature is $R=- d(d-1)\Lambda=-12\Lambda$. Note that our $\Lambda$ is related to the cosmological constant in its usual definition via $\Lambda_{cosm}=-6\Lambda$. $\Lambda > 0$ corresponds to $AdS$.

Note that this is actually $SO(3,2)$ covariant, as we can combine $\omega,e$ into a super-connection. Note that $\Lambda$ has units $(Length)^{-2}$.
In the $SO(3,2)$-invariant formalism, $I_{MM}$ arises from
\begin{equation}
I_{MM}=\frac{1}{2\Lambda}\int \epsilon_{ABCDE} V^E {\cal R}^{AB}\wedge{\cal R}^{CD}\,,
\end{equation}
where $V^E$ is a non-dynamical 0-form field (that we take to have value $V^{-1}=1$ to gauge back to the $SO(3,1)$ formalism) and ${{\cal R}^A}_B$ is the curvature of ${\Omega^A}_B\equiv \{ e^a,{\omega^a}_b\}.$ There are also quasi-topological terms of the form
\begin{equation}
I_{top}=\frac{\theta}{2\Lambda}\int {{\cal R}^A}_B\wedge {{\cal R}^B}_A+
\frac{\alpha}{\Lambda}\int {{\cal R}^A}_B\wedge {{\cal R}_{AC}}V^BV^C
\end{equation}
that we could add to the action.
In the stated gauge, this reduces to
\begin{equation}\label{eq:topaction}
I_{top}=\frac{\theta}{2\Lambda}P_2+(\theta+\alpha) C_{NY}+
\alpha\int R_{ab}\wedge e^a\wedge  e^b
\end{equation}
where $P_2=\int {R^a}_b\wedge {R^b}_a$ is the Pontryagin class, $C_{NY}=\int (T^a\wedge T_a-R_{ab}\wedge e^a\wedge  e^b)$ is the Nieh-Yan class and we also note the Euler class $E_2=\int\epsilon_{abcd} R^{ab}\wedge R^{cd}$. Note that in the presence of torsion, the action (\ref{eq:topaction}) contains the non-topological term $\int R_{ab}\wedge e^a\wedge  e^b$ with ``Immirzi parameter" $\gamma=-2/\alpha$. In the absence of torsion, this term is a total derivative. 

The Hilbert-Palatini action is
\beq
I_{HP}=I_{MM}-\frac{1}{2\Lambda}E_2\,.
\eeq
It differs from $I_{MM}$ by a boundary term,  is smooth as $\Lambda\to 0$ but is not manifestly $SO(3,2)$-invariant.  

\subsection{The $3+1$ split}
\newcommand\vt{{\bf t}}
\newcommand\vn{{\bf n}}
\newcommand\vN{{\bf N}}
\newcommand\ve{{\bf e}}
\newcommand\vvp{{\bf v}_\|}

Next, we carefully consider the $3+1$ split.   Although much of the discussion here is familiar from the ADM formalism, we feel it is important to set notation carefully, as we will introduce some new ingredients. To accommodate both $AdS$ and $dS$ signatures simultaneously, we will introduce a `time' function $t$ and a foliation of space-time $\Sigma_t\hookrightarrow M$. In $dS$, $t$ is time-like, and this corresponds to the usual Hamiltonian foliation; in $AdS$ on the other hand, we will take $t$ to be the (space-like) radial coordinate. We will keep track of the resulting signs by a parameter $\sigma_\perp$, equal to $\pm1$ in $dS$($AdS$). 

Proceeding as usual then, we get a vector field $\vt$ that satisfies $\nabla_\vt t=1\equiv \vt(t)$ (so $\vt=\frac{\partial}{\partial t}$) and  a 1-form $dt$. Given a 4-metric, we can introduce the normal 1-form $n$ as
\beq
 n=\sigma_\perp Ndt\,,
 \eeq
which is normalized as $(n,n)=\sigma_\perp$. The dual vector field $\vn$ can be expanded as
\beq
\vn=\frac{1}{N}\vt-\frac{1}{N}\vN\,,
\eeq
where the shift $\vN$ satisfies $(\vN,\vn)=0$, and thus $(\vt,\vn)=\sigma_\perp N$.

Next, we will locally choose a basis of 1-forms  
\beqn 
e^0=\sigma_\perp n=Ndt\,,\\
e^\alpha=\tilde e^\alpha+N^\alpha dt\,.
\eeqn
The $\tilde e^\alpha$ span $T^*\Sigma_t$, and correspond to a 3-metric $h_{ij}=\tilde e^\alpha_i\tilde e^\beta_j \eta_{\alpha\beta}$.
The quantities $N^\alpha$ are the components of $\vN$: $N^\alpha=e^\alpha_i N^i$.  These basis 1-forms are dual to $\{ \ve_0=\vn,\ve_\alpha=\tilde\ve_\alpha\}$, with $\ve_a(e^b)=\delta_a^b$. 

We expand the spin connection in the same 
basis\footnote{We have ${q^a}_b=N{{\omega_0}^a}_b+N^\alpha{{\omega_\alpha}^a}_b$ and ${{\tilde\omega}^a}_b \equiv {{\omega_\alpha}^a}_b{\tilde e}^\alpha$.}
\beq
{\omega^a}_b = {q^a}_b dt +{{\tilde\omega}^a}_{\kern5pt b}\,,
\eeq
which leads to 
\beq 
{R^a}_b={\tilde R^a}_{\kern5pt b}+dt\wedge {r^a}_b\,,
\eeq
where $\tilde R$ is formed from $\tilde\omega$ and $\tilde d$ only, and 
\beq
 {r^a}_b={\dot{\tilde\omega}^a}_{\kern5pt b}
 -{\tilde d}{q^a}_b
 -{\tilde\omega^a}_{\kern5pt c}{q^c}_b
+{q^a}_c{\tilde\omega^c}_{\kern5pt b}\,.
\eeq
Note that these quantities are merely decompositions along $T^*\Sigma_t$ in the 4-geometry; we will introduce the intrinsically defined objects shortly.

We then find
\begin{equation}
I_{HP}=2\epsilon_{\alpha\beta\gamma}\int dt\wedge\left\{ N({\tilde R}^{\alpha\beta}+\Lambda {\tilde e}^\alpha\wedge {\tilde e}^\beta)\wedge {\tilde e}^\gamma -2N^\alpha ( {\tilde R}^{0\beta})\wedge {\tilde e}^\gamma+r^{0\alpha}\wedge {\tilde e}^\beta\wedge {\tilde e}^\gamma\right\}\,.
\end{equation}
As is familiar, the lapse and shift appear as Lagrange multipliers. The constraints that they multiply are of course zero in any background (i.e. vacuum solution), such as (A)dS$_4$. The final term in the action contains the real dynamics -- $r^{0\alpha}$ depends on the components ${R^{0\alpha}}_{0\beta}$ of the Riemann tensor. 

Note though that the tensors used here are 4-dimensional. Let us define the "electric field"
\beq
K_\alpha = \sigma_\perp {\tilde\omega^0}\kern2pt_\alpha
=K_{\beta\alpha}\tilde e^\beta\,.
\eeq
In the case that $\omega$ is the torsion-free Levi-Civita connection, this agrees with the standard definition for extrinsic curvature, regarded as a vector-valued one-form. We then find
\begin{equation}\label{eq:GCone1}
\tilde R^{\alpha}_{\kern5pt\beta}=^{(3)}\kern-5pt{R^{\alpha}}_{\beta}- \sigma_\perp K^\alpha\wedge K_\beta\,,
\end{equation}
and
\begin{equation}\label{eq:GCtwo2}
{\tilde R^0}_{\kern5pt\alpha}=\sigma_\perp ( {\tilde {d}K}_\alpha+K_\beta\wedge\tilde\omega^\beta_{\kern5pt\alpha})\equiv\sigma_\perp (\tilde D K)_\alpha\,.
\end{equation}
These equations amount to the Gauss-Codazzi relations. 

Furthermore, $r^{0\alpha}$ contains time derivatives of ${\tilde\omega}^{0\alpha}$ as well as terms linear in components of $q$. We find
\begin{eqnarray}
2\epsilon_{\alpha\beta\gamma}r^{0\alpha}\wedge {\tilde e}^\beta\wedge {\tilde e}^\gamma &=&
 2\epsilon_{\alpha\beta\gamma}\left[ \sigma_\perp \dot K^\alpha-
 (\tilde Dq)^{0\alpha} \right]\wedge\tilde e^\beta\wedge\tilde e^\gamma, \\
&=&
 2\sigma_\perp\epsilon_{\alpha\beta\gamma}\left( \dot K^\alpha+ q^{\alpha\delta}K_\delta \right)\wedge\tilde e^\beta\wedge\tilde e^\gamma
+4q^{0\alpha}\left[\epsilon_{\alpha\beta\gamma}\tilde{T}^\beta\wedge\tilde{e}^\gamma\right]
\nonumber\end{eqnarray}
up to a total 3-derivative. We have defined the intrinsic 3-torsion $\tilde T^\alpha=\tilde d\tilde e^\alpha+{\tilde\omega^\alpha}\kern2pt_\beta\wedge \tilde e^\beta$.
Since we wish to regard the $\tilde e$ as coordinate variables,\footnote{Without this integration by parts, we would be in the Ashtekar formalism. Here, our choice gives a formalism closely related to the metric variable formalism. Note that the induced boundary term may be written $-\frac12 \Pi_\alpha\wedge \tilde e^\alpha$.} we integrate the first term by parts to obtain (up to the total time-derivative $\frac{\partial}{\partial t}\left( 2\sigma_\perp K^\alpha\wedge\tilde e^\beta\wedge\tilde e^\gamma\epsilon_{\alpha\beta\gamma}\right)$)
\beq
\label{r0a}
2\epsilon_{\alpha\beta\gamma}r^{0\alpha}\wedge {\tilde e}^\beta\wedge {\tilde e}^\gamma 
=\Pi_\alpha\wedge \dot{\tilde e}^\alpha
+ 4q^{0\alpha}\epsilon_{\alpha\beta\gamma}\tilde{T}^\beta\wedge\tilde{e}^\gamma+2\sigma_\perp q^{\alpha\delta}\epsilon_{\alpha\beta\gamma}K_\delta\wedge {\tilde e}^\beta\wedge {\tilde e}^\gamma\,.
\eeq
where we have defined the momentum 2-form 
\beq
\Pi_\alpha = -4\sigma_\perp \epsilon_{\alpha\beta\gamma} K^\beta\wedge\tilde e^\gamma\,.
\eeq
The $q^{ab}$ appear as Lagrange multipliers. In particular, the $q^{\alpha\beta}$ constraint precisely sets the antisymmetric (torsional) part of the extrinsic curvature tensor $K_{[\alpha\beta]}$ to zero. Next, we define the "magnetic field"
\beq
\label{defB}
B_\alpha=\frac12\sigma_\perp\epsilon_{\alpha\beta\gamma}\tilde\omega^{\beta\gamma},\ \ \ \ \omega^{\alpha\beta}=-\epsilon^{\alpha\beta\gamma}B_\gamma.
\eeq
and we find that the $q^{0\alpha}$ constraint 
\begin{equation}
\label{Antisymm}
\epsilon_{\alpha\beta\gamma}{\tilde T}^\beta\wedge {\tilde e}^\gamma=\epsilon_{\alpha\beta\gamma}\tilde{d}\tilde{e}^\beta\wedge\tilde{e}^\gamma -\sigma_\perp B_{\beta}\wedge\tilde{e}^\beta\wedge\tilde{e}_\alpha=0\,,\end{equation}
involves only the antisymmetric part $B_{[\alpha,\beta]}$ of  the magnetic field $B_\alpha=B_{\alpha\beta}\tilde{e}^\beta$.  The antisymmetric part of $B_\alpha$ spoils the gauge covariance of the constraint (\ref{Antisymm}) under an SO(3) rotation of the dreibein $\tilde{e}^\alpha$, hence it represents degrees of freedom that can be gauged fixed to zero by an SO(3) rotation. On the other hand, an algebraic equation of motion connects the symmetric part of $B_{\alpha\beta}$ to derivatives of $\tilde{e}^\alpha$ as
 \begin{equation}
 \label{deB}
\tilde{d}\tilde{e}^\alpha +\epsilon^{\alpha\beta\gamma}B_\beta\wedge\tilde{e}_\gamma=0
\end{equation}
At the end, one is left with the canonically conjugate variables $\tilde{e}^\alpha$ and $\Pi_\alpha$.  These results are familiar from the metric formalism.

Dropping the torsional terms, we then arrive at the action
\beqn
I_{HP}&=&\int dt\wedge \Bigl\{ \dot{\tilde e}^\alpha\wedge \Pi_\alpha 
+ 2N\epsilon_{\alpha\beta\gamma}( ^{(3)}R^{\alpha\beta}-\sigma_\perp K^\alpha\wedge K^\beta+\Lambda {\tilde e}^\alpha\wedge {\tilde e}^\beta)\wedge {\tilde e}^\gamma \nonumber \\
 && -4\sigma_\perp N^\alpha\epsilon_{\alpha\beta\gamma}(\tilde D K)^\beta\wedge {\tilde e}^\gamma\Bigl\}\,.
\eeqn
Furthermore, using $*_3 \tilde e^\alpha\wedge\tilde e^\beta = \frac12\eta_{\gamma\delta}\epsilon^{\alpha\beta\gamma}\tilde e^\delta$, we have
\beq
\hat\Pi_\alpha =*_3\Pi_\alpha=-2(K_{\alpha\beta}-\eta_{\alpha\beta}tr K ) \tilde e^\beta \,,
\eeq
where $tr K=\eta^{\alpha\beta}K_{\alpha\beta}$. 
\comment{This is strictly correct. Writing the RHS in terms of $K_\alpha$ requires torsional part to vanish.}
We can solve the above equation to get
\beq
K_\alpha = -\frac{1}{2}(\hat\Pi_{\alpha\beta}-\frac12 \eta_{\alpha\beta}tr\hat\Pi) \tilde e^\beta\,.
\eeq
As stated above, $K_{\alpha\beta}$ (and $\hat\Pi_{\alpha\beta}$) is symmetric when the torsion vanishes.

Finally, with the definition (\ref{defB}) we find\footnote{The spatial signature $\sigma_3$ appears in $\epsilon_{\alpha\beta\gamma}\epsilon_{\phi\delta\rho}\eta^{\alpha\phi}=\sigma_3(\eta_{\beta\delta}\eta_{\gamma\rho}-\eta_{\beta\rho}\eta_{\gamma\delta})$. We will always consider Lorentzian spacetime signature, so $\sigma_3=-\sigma_\perp$.}
\begin{eqnarray}
{\epsilon_{\alpha\beta\gamma}}^{(3)}R^{\alpha\beta}\wedge \tilde e^\gamma =
\epsilon_{\alpha\beta\gamma}\left[ \tilde d {\tilde\omega}^{\alpha\beta}+{\tilde\omega^\alpha}_{\kern5pt\delta}\wedge {\tilde\omega^{\delta\beta}}\right]\wedge\tilde e^\gamma=
\sigma_\perp\left[ 2\tilde dB_\gamma+\epsilon_{\alpha\beta\gamma}B^\alpha\wedge B^\beta\right]\wedge\tilde e^\gamma\,.
\end{eqnarray}

Introducing $B_\alpha$ is an unusual thing to do but it will play a role in  duality: in this form, the Hamiltonian contains terms which are reminiscent of those of the Maxwell theory. The full HP action is of the form\beqn\label{HP_action}
I_{HP}&=&\int dt\wedge\left\{ \dot{\tilde e}^\alpha\wedge \Pi_\alpha -4\sigma_\perp N^\alpha\epsilon_{\alpha\beta\gamma}(\tilde D K)^\beta\wedge {\tilde e}^\gamma\right.\nonumber  \\
&&\left. + 2\sigma_\perp N(2\tilde dB_\gamma + \epsilon_{\alpha\beta\gamma}B^\alpha\wedge B^\beta- \epsilon_{\alpha\beta\gamma}K^\alpha\wedge K^\beta+\sigma_\perp\Lambda \epsilon_{\alpha\beta\gamma}\tilde e^\alpha\wedge \tilde e^\beta)\wedge\tilde e^\gamma\right\}\,.
\eeqn
Note that the entire contribution of the cosmological constant appears in the last term of the Hamiltonian constraint.

\subsection{Shifted Variables}

It is possible to make a transformation of the canonical variables in order to absorb the cosmological constant term in (\ref{HP_action}). This can be achieved by introducing the new variables
\beq\label{dualvar}\hat K^\alpha = K^\alpha-\rho \tilde e^\alpha\,,
\eeq
and requiring that
\beq
\rho^2 =\sigma_\perp \Lambda\,.
\eeq
This is positive only when $\sigma_\perp$ and $\Lambda$ are simultaneously positive or negative, as it is the case for both AdS$_4$ ($\Lambda >0$) and dS$_4$ ($\Lambda <0$). We will often write $\Lambda=\sigma_\perp /L^2$ where $L$ is a length scale.

Under (\ref{dualvar}) the momentum 2-form becomes 
\beq
\label{PiTransf}
\Pi_\alpha \rightarrow {\cal P}_\alpha -4\sigma_\perp\rho\epsilon_{\alpha\beta\gamma}\tilde e^\beta\wedge\tilde e^\gamma\,.
\eeq
The last term in (\ref{PiTransf}) contributes a total time derivative to the action (of the form of a boundary cosmological term). We have introduced
a new momentum variable
\[ 
{\cal P}_\alpha = -4\sigma_\perp \epsilon_{\alpha\beta\gamma} \hat K^\beta\wedge\tilde e^\gamma \,.
\]
Then, we get the action
\beqn
I_{HP}=\int dt\wedge\Bigl\{ \dot{\tilde e}^\alpha\wedge {\cal P}_\alpha -4\sigma_\perp N^\alpha\epsilon_{\alpha\beta\gamma}(\tilde D \hat K+\rho \tilde T)^\beta\wedge {\tilde e}^\gamma 
-\frac43\sigma_\perp\rho\epsilon_{\alpha\beta\gamma}\frac{\partial}{\partial t}({\tilde e}^\alpha\wedge {\tilde e}^\beta\wedge {\tilde e}^\gamma)
\nonumber \\
+ 2\sigma_\perp N\left(2\tilde d (B_\alpha\wedge\tilde e^\alpha)+2B_\gamma\wedge\tilde T^\gamma- \epsilon_{\alpha\beta\gamma}\left(
B^\alpha\wedge B^\beta+\hat K^\alpha\wedge \hat K^\beta
+ 2\rho\hat K^\alpha\wedge \tilde e^\beta
\right)\wedge\tilde e^\gamma\right)\Bigl\}\,.
\eeqn
Note that the shift constraint is still written in terms of the ordinary covariant derivative, and thus involves a non-linear term coupling $B$ to $\hat K$. Consistent with our previous discussion, we drop the terms involving the torsion $\tilde T$, and disregard the boundary term to obtain
\beqn
\label{IHPfull}
I_{HP}&=&\int dt\wedge\Bigl\{ \dot{\tilde e}^\alpha\wedge {\cal P}_\alpha -4\sigma_\perp N^\alpha\epsilon_{\alpha\beta\gamma}(\tilde D \hat K)^\beta\wedge {\tilde e}^\gamma 
\nonumber \\
&&+ 2\sigma_\perp N\left(2\tilde d (B_\alpha\wedge\tilde e^\alpha)
- \epsilon_{\alpha\beta\gamma}\left(
B^\alpha\wedge B^\beta+\hat K^\alpha\wedge \hat K^\beta
+ 2\rho\hat K^\alpha\wedge \tilde e^\beta
\right)\wedge\tilde e^\gamma\right)\Bigl\}\,.
\eeqn
We note that the parameter $\rho$ can be of {\it either} sign (although, this sign does not appear in the second order equations of motion).

\subsection{Linearization}

Next, we linearize the above action around an appropriate fixed background. We expand as
\beq
\label{background}
\tilde{e}^\alpha =\underline{\tilde{e}}^\alpha +E^\alpha,\,\,N=1+n,\,\,\,\, N^\alpha =n^\alpha, \,\,\,B^\alpha =\underline{B}^\alpha +b^\alpha,\,\,\, \hat {K}^\alpha =\underline{\hat K}^\alpha +k^\alpha\,.
\eeq
The background values should satisfy the constraints. The simplest choice is the background where
\beq\label{adsbg}
\underline{\hat K}^\alpha =0=\underline{B}^\alpha\,.
\eeq
In fact, reaching this simple form was a motivation for the shift (\ref{dualvar}).
Then, to quadratic order in the fluctuating fields the Hamiltonian gives
\beqn
\label{linHPfinal}
I_{HP}=\int dt\wedge\left\{ \dot E^\alpha\wedge p_\alpha -4\sigma_\perp n^\alpha\epsilon_{\alpha\beta\gamma}\tilde d k^\beta\wedge \underline{\tilde e}^\gamma 
+4\sigma_\perp n\left( \tilde d (b_\alpha\wedge \underline{\tilde e}^\alpha)-\rho\epsilon_{\alpha\beta\gamma} k^\alpha\wedge \underline{\tilde e}^\beta\wedge\underline{\tilde e}^\gamma\right)
\right.\nonumber \\
\left. - 2\sigma_\perp \epsilon_{\alpha\beta\gamma}\left(
b^\alpha\wedge b^\beta+k^\alpha\wedge k^\beta
+ 2\rho k^\alpha\wedge E^\beta
\right)\wedge\underline{\tilde e}^\gamma\right\}\,,
\eeqn
where 
\beq
p_\alpha = -4\sigma_\perp \epsilon_{\alpha\beta\gamma} k^\beta\wedge\underline{\tilde e}^\gamma
\eeq
are the linearized momentum variables conjugate to $E^\alpha$. 

In order to reach the form (\ref{linHPfinal}) the linear terms in the fluctuations must vanish. For this to happen we find the relationships
\beq\label{edotpm}
\dot{\underline{\tilde{e}}}^\alpha+\rho \underline{\tilde{e}}^\alpha=0\,.
\eeq 
Notice that we can also write the linearized action in the form
\beqn
I_{HP}=\int dt\wedge\left\{ (\dot E^\alpha+\rho E^\alpha)\wedge p_\alpha- 2\sigma_\perp \epsilon_{\alpha\beta\gamma}\left(
b^\alpha\wedge b^\beta+k^\alpha\wedge k^\beta
\right)\wedge\underline{\tilde e}^\gamma\right.\nonumber 
\\
\left.
 -4\sigma_\perp n^\alpha\epsilon_{\alpha\beta\gamma}\tilde d k^\beta\wedge \underline{\tilde e}^\gamma 
+n\left( 4\sigma_\perp \tilde d b_\gamma+\rho p_\gamma\right)\wedge\underline{\tilde e}^\gamma
 \right\}\,.
\eeqn
The form of the first term, involving the momentum, makes clear that longitudinal fluctuations 
are non-dynamical. The natural time dependence of $E^\alpha$ is of the form $e^{-\rho t}$ (correspondingly, the natural time dependence of $p_\alpha$ is $e^{+\rho t}$).
Other than that, we see that in comparing to the flat space action, in these variables, the only change is that the Hamiltonian constraint is modified.

The solutions of (\ref{edotpm}) and (\ref{adsbg}) are components of (A)dS$_4$ spacetimes. We can solve (\ref{edotpm}) to obtain
\beq
\underline{e}^0=dt,\ \ \underline{e}^\alpha = e^{-\rho t}dx^\alpha\,.
\eeq
With these we construct the usual Poincar\'e metric on  (A)dS which, however,  covers only half of the space even though the parameter $t$ runs from $-\infty$ to $+\infty$.  The conformal boundary in these coordinates is at $t=+\infty$.
Then we derive
\beq
{\underline{\omega}^\alpha}_{0}=-\rho e^{-\rho t}dx^\alpha=-\rho\underline{e}^\alpha\,,
\eeq
and so
\beqn
\left.
\begin{array}{ll}
 {\underline{R}^\alpha}_\beta = 
 -\frac{\sigma_\perp}{L^2}\underline{e}^\alpha\wedge \underline{e}_\beta\\
 {\underline{R}^\alpha}_0 = 
 -\frac{1}{L^2}\underline{e}^\alpha\wedge \underline{e}^0   
\end{array}
\right\}
{\underline{R}^a}_b = 
-\frac{\sigma_\perp}{L^2} \underline{e}^a\wedge \underline{e}_b\,.
\eeqn
Hence $\underline{Ric}_{ab}=-\frac{3\sigma_\perp}{L^2}\eta_{ab}$ and $\underline{R}=-12\sigma_\perp/L^2=-12\Lambda$. We also evaluate
\beqn
\underline{\Pi}_\alpha = -4\sigma_\perp\rho\epsilon_{\alpha\beta\gamma}\underline{{\tilde e}}^\beta\wedge\underline{{\tilde e}}^\gamma,\ \ \ \
\underline{\hat\Pi}^\alpha&=&4\rho
\underline{{\tilde e}}^\alpha,\ \ \ tr\hat{\underline{\Pi}}=12\rho\\
\underline{B}_\alpha = 0,\ \ \ && \ \
\label{K+} 
\underline{K}^\alpha = \rho \underline{\tilde e}^\alpha\Rightarrow \hat {\underline{K}}^\alpha =0
\eeqn
Note that in this gauge, $(\underline{\tilde D} \underline{K})^\alpha =\frac1L \underline{\tilde T}^\alpha=0$, which solves the shift constraint, while the Hamiltonian constraint is satisfied through a cancellation between the $\underline{K}^2$ term and the cosmological term.

\section{Linearized Gravitational Duality and Holography}

Let us summarize what we have obtained so far. In the presence of a cosmological constant we have defined variables such that the action resembles most closely the action without the cosmological constant. This was done in order to look for a suitable background around which linear fluctuations are as simple as possible. Requiring that $\hat K$ (the ``electric field") and $B$ (the "magnetic field") vanish in such a background - as they do around flat space - we found that the background should be (A)dS$_4$. 
Quite satisfactorily, {\it both} sign choices for $\rho$ in the change of variables 
(\ref{dualvar}) lead to (A)dS$_4$ spacetimes. 

\subsection{Duality and Holography}

This is the appropriate point to recall some salient features of duality rotations. In simple Hamiltonian systems 
the effect of the canonical transformation $p\mapsto q$ and $q\mapsto -p$  to the action is (see e.g. \cite{Goldstein})
\beq
\label{duality3}
I=\int_{t_1}^{t_2} dt[p\dot{q}-H(p,q)]\mapsto I_D=\int_{t_1}^{t_2}dt [-q\dot{p}-H(q,-p)]\,.
\eeq
Notice that $I_D$ involves the {\it dual} variables, for which we have however kept the same notation for simplicity. The transformed Hamiltonian $H(q,-p)$ is in general not related to $H(p,q)$. However, if
$
H(q,-p) =H(p,q)$ we call the above transformation a {\it duality}. It then holds
\beq
\label{duality44}
I_D =I-qp\Bigl|^{t_2}_{t_1}\,.
\eeq
The dual action describes exactly the same dynamics as the initial one, up to a modification of the boundary conditions. For example, if $I$ is stationary on the e.o.m for fixed $q$ in the boundary, $I_D$ is stationary on the same e.o.m. for fixed $p$ in the boundary. This simple example illustrates the role of duality in holography; a bulk duality transformation  corresponds to a particular modification  of the boundary conditions. This property of duality transformations is behind the remarkable holographic properties of electormagnetism  in (A)dS$_4$ \cite{Witten_sl2z,LP1}.

Clearly, the crucial properties of a duality transformation are to be canonical and to leave the Hamiltonian unchanged.   However, consider a slight generalization
\beq
\label{duality5}
S=\int_{t_1}^{t_2}dt[p\dot q-\frac{1}{2}(p^2+q^2+2\lambda pq)]
\eeq
where $\lambda$ is an arbitrary parameter.  
The Hamiltonian now is {\it not} invariant under the canonical transformation $p\mapsto q$ and $q\mapsto -p$ -- the $pq$ term changes sign. Consequently, the first order form of the equations of motion are also not duality invariant. Nevertheless, the second order equation of motion is invariant. We will find that gravity in the presence of a cosmological constant follows precisely this model. Of course, gravity is a much more complicated constrained system, but as we will show, the constraints and Bianchi identities transform appropriately.

We also note that the canonical transformation (implemented by a generating functional of the first kind)
\beq
\label{duality6}
p\mapsto q+2\lambda p\,,\,\,\,\, q\mapsto -p\,.
\eeq
is of interest here.
The above does not change the Hamiltonian and the transformed action differs from the initial one by total time derivative terms\footnote{In holography, the latter terms correspond to the relevant "multi-trace"  boundary deformations discussed  in \cite{LP1}.} 
\beq
\label{duality7}
S\mapsto S_D=S -pq\Bigl|^{t_2}_{t_1} -\lambda p^2 \Bigl|^{t_2}_{t_1}\,.
\eeq

\subsection{Linearized gravitational duality}

As a preamble to gravity we recall  the duality properties of Maxwell theory 
\beq
\label{Maxw1}
I_{Max}=\frac{1}{2g^2}\int dt\wedge\left\{ \dot{\tilde A}\wedge *_3 E-\frac{1}{2}(E\wedge *_3E+B\wedge *_3B)-A_0\tilde d *_3E\right\}\,,
\eeq
Under the duality $E\mapsto -*_3B$, $B\mapsto *_3E$, $\tilde A\mapsto \tilde A_D$, we find
\beq
\label{Maxw2}
I_{Max}\mapsto I_{Max,D}=\frac{1}{2g^2}\int dt\wedge\left\{ -\dot{\tilde A}_D\wedge B-\frac{1}{2}(E\wedge *_3E+B\wedge *_3B)+A_{0}\tilde d B\right\}\,.
\eeq
$E$ and $B$ in (\ref{Maxw2}) should be expressed through $\tilde{A}_D$. We observe that the kinetic term has changed sign, while the Hamiltonian remains invariant. In addition, the (Gauss) constraint is dualized to the trivial `Bianchi' identity $dB=0$ for the dual magnetic field.

Next we try to apply a Maxwell-type duality map in gravity. We consider the following transformation around the fixed background (\ref{edotpm})
\beq
\label{dualitymap}
k^\alpha\mapsto - b^\alpha,\ \ \ \ \ \ \ b^\alpha\mapsto k^\alpha\,.
\eeq
To implement the map (\ref{dualitymap}) we need to specify the mapping
of $E^\alpha$ to   a `dual 3-bein' ${\cal E}^\alpha$. We do that using the linearized form of (\ref{deB}) as 
\beq
\label{calE}
\epsilon^{\alpha\beta\gamma}b_\beta\wedge \underline{\tilde e}_\gamma+\tilde d E^\alpha=0
\mapsto 
\epsilon^{\alpha\beta\gamma}k_{\beta}\wedge \underline{\tilde e}_\gamma+\tilde d {\cal E}^\alpha=0=\tilde d{\cal E}^\alpha-\frac{1}{4\sigma_\perp}p_\alpha
\eeq
Since $p_\alpha = 4\sigma_\perp \tilde d {\cal E}_\alpha$, it is natural to define 
\beq
\label{pDual}
p_{D,\alpha}=4\sigma_\perp \tilde d E_\alpha = -4\sigma_\perp \epsilon_{\alpha\beta\gamma}b^\beta\wedge \underline{\tilde e}^\gamma\,,
\eeq
 and thus the mapping (\ref{dualitymap}) is supplemented by
\beq
E\mapsto {\cal E}\,,\,\,\, {\cal E} \mapsto -E\,,\,\,\,
p\mapsto -p_D\,,\,\,\, p_D \mapsto p
\eeq

Now, let us see the effects of the above duality mapping. The action transforms to
\beqn
\label{IHPD}
I_{HP}\mapsto I_{HP,D}=\int dt\wedge\left\{ -\dot {\cal E}^\alpha\wedge p_{D,\alpha}-\rho {\cal E}^\alpha\wedge p_{D,\alpha}- 2\sigma_\perp \epsilon_{\alpha\beta\gamma}\left(
b^\alpha\wedge b^\beta+k^\alpha\wedge k^\beta
\right)\wedge\underline{\tilde e}^\gamma\right.\\
\left.
 +4\sigma_\perp n^\alpha\epsilon_{\alpha\beta\gamma}\tilde d b^\beta\wedge \underline{\tilde e}^\gamma 
+n\left( 4\sigma_\perp \tilde d k_\gamma+\rho p_{D,\alpha}\right)\wedge\underline{\tilde e}^\gamma
 \right\}\nonumber\,
\eeqn
where now $k^\alpha$ and $b^\alpha$ should be expressed in terms of the dual variables ${\cal E}^\alpha$ and $p_{D,\alpha}$ via (\ref{calE}) and (\ref{pDual}). We notice that the 'kinetic' part $\dot{{\cal E}}\wedge p$  of the action changes sign under the duality map, in direct analogy with the Maxwell case. However, the  Hamiltonian is not invariant due to the change of sign of the second term in the first line of (\ref{IHPD}). We will discuss this further in a later section. For now, we note that this sign change would not show up in the equations of motion, {\it written in second order form.} It is important to also  note that the constraints are transformed into quantities which in the next subsection we will recognize as the linearized Bianchi identities. This is to be expected since the duality transformations are canonical. We also note that it may be possible to choose an alternative canonical transformation, designed to leave the Hamiltonian invariant. The latter is presumably related to the work of Julia et. al. \cite{Julia} and is considered in the Appendix.
 
 \subsection{Linearized Constraints and Bianchi Identities}

By virtue of the discussion above we may now demonstrate that under the duality mapping (\ref{dualitymap}) the linearized constraints transform to the linearized Bianchi identities as
\beqn
\label{ConBianchi1}
C_\alpha\equiv\epsilon_{\alpha\beta\gamma}\tilde d k ^\beta\wedge \underline{\tilde e}^\gamma
&\mapsto&
-\epsilon_{\alpha\beta\gamma}\tilde d b^\beta\wedge \underline{\tilde e}^\gamma
\\
\label{ConBianchi2}
C_0\equiv-\sigma_\perp\left( \tilde d b_\gamma - \rho \epsilon_{\alpha\beta\gamma} k ^\alpha\wedge \underline{\tilde e}^\beta\right)\wedge\underline{\tilde e}^\gamma
&\mapsto&
-\sigma_\perp\left( \tilde d k_{\gamma}+ \rho \epsilon_{\alpha\beta\gamma} b^\alpha\wedge \underline{\tilde e}^\beta\right)\wedge\underline{\tilde e}^\gamma
\eeqn
To identify the right hand sides, we first note that the Bianchi identities are 
\beqn
{{B_R}^a}_b &=& d{R^a}_b - {R^a}_c\wedge {\omega^c}_b+{\omega^a}_c\wedge {R^c}_b=0\\
B_T^a &=& d{T^a} - {R^a}_b\wedge e^b+{\omega^a}_b\wedge T^b=0
\eeqn
which are obtained from the definitions of ${R^a}_b$ and $T^a$ by exterior differentiation. The first equation is satisfied identically. 
Since the torsion vanishes, the second equation tells us only that ${R^a}_b\wedge e^b=0$. If we do the 3+1 split, we find two equations. The first is
\beq
{B_T}^\alpha=-(^{(3)}{R^\alpha}_\beta-\sigma_\perp K^\alpha\wedge K_\beta)\wedge \tilde e^\beta=0
\eeq
which upon using the symmetry of $K^\alpha$ linearizes to 
\beq
{B_T}^\alpha=-\epsilon_{\alpha\beta\gamma}\tilde d b^\beta\wedge \underline{\tilde e}^\gamma+\ldots
\eeq
Note that this is the image under duality of the shift constraint as in (\ref{ConBianchi1}). 

The second identity is
\beqn {B_T}^0&=&-{\tilde R}^0\kern.2pt_\alpha\wedge\tilde e^\alpha =-\sigma_\perp (\tilde D K)_\alpha\nn\\
&=&-\sigma_\perp\left(  \tilde dk_\alpha+\rho\epsilon_{\alpha\beta\gamma}b^\beta\wedge \underline{\tilde e}^\gamma\right)\wedge\underline{\tilde e}^\alpha=0
\eeqn
where to arrive in the second line we used (\ref{K+}).  This is the image of the Hamiltonian constraint as in (\ref{ConBianchi2}).

Summarizing, the duality transformations between linearized constraints and Bianchi identities are
\beqn
C_\alpha\mapsto B_{T,\alpha} && C_0\mapsto  B_T^{0}\\
B_{T,\alpha}\mapsto -C_\alpha && B_T^{0}\mapsto -C_0
\eeqn

\subsection{Connection with other known dualities}

The Maxwell-type duality operation (\ref{dualitymap}) is closely related to the dualization of the first two indices of the Riemann tensor as\footnote{For a discussion of the duality properties of gravity in terms of the Riemann tensor see \cite{Hull}.}
\beq\label{Riemdual}
{R^a}_b\to {S^a}_b\equiv \frac12{{\epsilon^a}_{bc }}^d {R^c}_d
\eeq
at least at the linearized level. 
Let us investigate (\ref{Riemdual}) by rewriting expressions in the 3+1 split. We have
\[ {R^a}_b={\tilde R^a}_{\kern5pt b}+dt\wedge {r^a}_b\]
\[ {S^a}_b={\tilde S^a}_{\kern5pt b}+dt\wedge {s^a}_b\]
We begin with the spatial 2-forms when we have
\begin{equation}\label{eq:GCone}
\tilde R^{\alpha\beta}=-\epsilon^{\alpha\beta\gamma}\tilde d B_\gamma+ \sigma_\perp (B^\alpha\wedge B^\beta-K^\alpha\wedge K^\beta)
\end{equation}
\begin{equation}\label{eq:GCtwo}
{\tilde R^0}_{\kern5pt\alpha}=\sigma_\perp ( {\tilde {d}K}_\alpha+K_\beta\wedge\tilde\omega^\beta_{\kern5pt\alpha})\equiv\sigma_\perp (\tilde D K)_\alpha
\end{equation}
and
\beqn
{\tilde{S}^0}\kern0.2pt_\gamma &=& \frac12\sigma_\perp\epsilon_{\alpha\beta\gamma}\tilde{R}^{\alpha\beta}\\
{\tilde{S}_{\alpha\beta }} &=&\epsilon_{\alpha\beta\gamma}\tilde{R}^{0\gamma}\eeqn
If we linearize these expressions, we find under the duality transformation (\ref{dualitymap})
\beq
\tilde R^{ab}\mapsto -\sigma_\perp \tilde S^{ab}
\eeq
Because the expressions (\ref{eq:GCtwo}) involve derivatives of $B$ and $K$, the duality (\ref{dualitymap}) is an `integrated form' of the usual Riemann tensor duality, but implies it.

Similarly, if we investigate the spatial 1-forms, we find
\beq
r^{ab}\mapsto -\sigma_\perp s^{ab}
\eeq
To arrive at this result we have set to zero the Lagrange multiplier field $q$.

\section{The Effect on the Boundary Theory}

It is well known that AdS is holographic. We may well ask, in the context of AdS/CFT, how the duality transformation that we have defined here acts in the boundary. We are instructed to consider the on-shell bulk action as a function of bulk fields. So, we evaluate the action on a solution to the equation of motion, resulting in a pure boundary term which is of the form
\beq
S_{bdy} = \int_{\partial M} p_\alpha\wedge E^\alpha
\eeq
Applying the duality transformation to the bulk theory, although the bulk action is not invariant as we have discussed above, nevertheless it may be easily shown that it induces a simple transformation on the (linearized) boundary term: it simply changes its sign.
\beq
S^{dual}_{bdy} = -\int_{\partial M} p_{D,\alpha}\wedge {\cal E}^\alpha
\eeq
This transformation is exactly analogous to what happens in the Maxwell case: it amounts to the result \cite{Tassos}.\footnote{See also \cite{Sachdev} for an interesting recent application of this formula.} 
\beq
G_2 G_2^{dual} = -1\,.
\eeq

\section{Conclusions and Outlook}

Motivated by possible application in holography and in higher-spin gauge theory we have studied the duality properties of gravity in the Hamiltonian formulation. We have presented the gravity action in terms of suitable variables that closely resemble the electric and magnetic fields in Maxwell theory. We have found suitable "electric" and "magnetic" field variables, such that at the linearized level first order gravity most closely resembles electromagnetism. This can be done {\it only} around Minkowksi and (A)dS$_4$ backgrounds. 

We have implemented duality transformations in the linearized gravity fluctuations around these backgrounds. In the presence of a cosmological constant, the Hamiltonian changes, nevertheless the bulk dynamics remains unaltered, while the linearized lapse and shift constraints are mapped into the linearized Bianchi identities. 
Moreover, the duality transformations induce boundary terms whose relevance in holography we have briefly discussed. Finally, we have exhibited a modified duality rotation that leaves the bulk Hamiltonian invariant, while it induces boundary terms corresponding to relevant deformations. 

The main implication of our results is that certain properties of correlations functions in three-dimensional CFTs mimic the duality of gravity. It would be interesting to extend our results to black-hole backgrounds and also when topological terms are present in the bulk. We also expect that one can analyze the duality of higher-spin gauge theories based on our first-order approach. 

{\bf Acknowledgments} 

The work of A. C. P. was partially supported by the research program "PYTHAGORAS II" of the Greek Ministry of Education. {\small RGL} was supported
in part by the U.S. Department of Energy under contract
DE-FG02-91ER40709.

\section{Appendix: other duality mappings}

It is possible to find a transformation that leaves the Hamiltonian unchanged. Consider the following transformation in the fixed background (\ref{edotpm})
\beq
\label{dualitymapNEW}
k^\alpha\mapsto - b^\alpha -2\rho {\cal E}^\alpha,\ \ \ \ \ \ \ b^\alpha\mapsto k^\alpha\,.
\eeq
The mapping to the dual dreibein is still specified by (\ref{calE}). 
A straightforward calculation reveals that the action transforms as
\beqn
\label{IHPDnew}
I_{HP}\mapsto I_{HP,D}&=&I_{HP}+4\sigma_\perp\int_{\partial M} \left(\epsilon_{\alpha\beta\gamma}{\cal E}^\alpha\wedge b^\beta+\rho \epsilon_{\alpha\beta\gamma}{\cal E}^\alpha\wedge{\cal E}^\beta\right)\wedge \underline{\tilde e}^\gamma\nonumber \\
 && +\int dt\wedge \Bigl[ 4\sigma_\perp n^\alpha \epsilon_{\alpha\beta\gamma}\tilde{d}b^\beta\wedge\underline{\tilde e}^\gamma-8\rho n^\alpha k_{\beta}\wedge \underline{\tilde e}_{\alpha}\wedge\underline{\tilde e}^\beta \nonumber \\
  && \hspace{.5cm}+n(4\sigma_\perp\tilde{d}k_\alpha +4\sigma_\perp\rho \epsilon_{\alpha\beta\gamma}b^\beta\wedge\underline{\tilde e}^\gamma +8\Lambda\epsilon_{\alpha\beta\gamma}{\cal E}^\beta\wedge\underline{\tilde e}^\gamma)\wedge\underline{\tilde e}^\alpha\Bigl]
  \eeqn
The transformations (\ref{dualitymapNEW}) leaves unchanged the Hamiltonian and changes the action by the total "time" derivative terms shown in the first line of (\ref{IHPDnew}).  

Moreover, the linearized constraints transform into the linearized Bianchi identities. Let us see that in some detail. The second term in the shift constraint is  zero since $k_\alpha$ is a symmetric one form $k_\alpha=k_{\alpha\beta}\underline{\tilde e}^\beta$ with $k_{\alpha\beta}=k_{\beta\alpha}$; see (\ref{r0a}). 

The term proportional to $\Lambda$ in the lapse constraint is also zero. This is slightly  more involved to see and it is based on the possibility of solving (\ref{calE}) for ${\cal E}^\alpha$ after gauge fixing.\footnote{This is the equivalent of inverting $\bar{E}=\nabla\times\bar{A}$ in the discussion of duality in electromagnetism \cite{DT}.} One way to see this is in components. Write ${\cal E}^\alpha ={\cal E}^\alpha_{\,\,\beta}\underline{\tilde e}^\beta$ and (\ref{calE}) becomes
\beq
\partial_\alpha{\cal E}^\beta_{\,\,\gamma} -\partial_{\gamma}{\cal E}^\beta_{\,\,\alpha} =\epsilon^\beta_{\,\,\,\delta\alpha}k^{\delta}_{\,\,\gamma} -\epsilon^\beta_{\,\,\,\delta\gamma}k^{\delta}_{\,\,\alpha}
\eeq
In the "Lorentz gauge" where $\partial^\alpha{\cal E}^\beta_{\,\,\alpha} =0 =\partial^\alpha k^\beta_{\,\,\alpha}$ the above can be inverted as
\beq
\label{EInverted}
{\cal E}^\alpha_{\,\,\beta} =\frac{1}{\partial^2}\epsilon^\alpha_{\,\,\delta\gamma}\partial^\gamma k^\delta_{\,\,\beta}
\eeq
Using (\ref{EInverted}) one verifies that the last term in the lapse constraint vanishes. This modified duality transformation is probably related to the one considered by Julia et. al. in \cite{Julia}.

\end{document}